\documentclass[acmsmall,nonacm]{acmart}
\usepackage{algorithmic}
\usepackage{graphicx}
\usepackage{textcomp}
\usepackage{xcolor}
\usepackage[skins]{tcolorbox}
 \usepackage{booktabs}
 \usepackage{tabularx}
\usepackage{url}
\DisableLigatures[-]{family=tt*}

\begin{document}

\newcommand{\rqone}{\textit{How common are oversight-loosening directives in agent skills, and through which mechanisms do they loosen oversight?}}
\newcommand{\rqtwo}{\textit{What kinds of skills are more likely to contain loosening directives?}}
\newcommand{\rqthree}{\textit{How do loosening directives propagate as skills are copied and edited?}}
\newcommand{\rqfour}{\textit{How well do existing skill scanners detect loosening directives?}}

\title{When Skills Tell Agents Not to Ask: A Large-Scale Study of Oversight Directives in Agent Skills}

\author{Timofey Sanko}
\affiliation{%
  \department{School of Computing}
  \institution{Queen's University}
  \city{Kingston}
  \state{Ontario}
  \country{Canada}}
\email{timofey.sanko@queensu.ca}

\author{Yuan Tian}
\affiliation{%
  \department{School of Computing}
  \institution{Queen's University}
  \city{Kingston}
  \state{Ontario}
  \country{Canada}}
\email{y.tian@queensu.ca}

\begin{abstract}
Agent skills, reusable instruction packages centred on a \texttt{SKILL.md} file, extend AI agents on demand, and a single line in a widely copied skill reaches every agent that loads it. Some lines are \emph{oversight-loosening directives}: they tell the agent to proceed without approval, to bypass checks with \texttt{--no-verify}, or to keep working without checking in. Although such directives can remove a checkpoint from every agent that loads them, no prior work has examined them. We present the first large-scale study of oversight-loosening directives, analyzing 1.53 million English skills from GitSkills. We split skills into units (sentences, bullets, table cells, and code lines), develop a codebook that labels oversight directives by polarity, mechanism, and scope, label 2,097 units by hand, and train a detector to scale the analysis. Our results show that about one in seventeen skills (about 88,000 skills) contains a loosening directive. These directives most often loosen oversight through permission flags, and most of them (70.1\%) apply unconditionally rather than only in specific contexts such as CI. Compared with skills that have the same amount of oversight language, skills that run shell code, install packages, or use \texttt{sudo} have 1.7--1.8 times the odds of containing a loosening directive, an association that holds across all our robustness checks, and they do not compensate with more tightening. Loosening spreads through a few upstream skills rather than through edits by their adopters: 2.4\% of the skill families that carry loosening account for half of all copies of loosening-carrying skills, and 1.1\% of distinct loosening units for half of all loosening occurrences. Open-source skill scanners catch at most 46\% of loosening directives, with a precision below 0.4, while the LLM behind them, prompted with our codebook instead, catches 61\% at a precision of 0.71. 

\end{abstract}

\maketitle

\section{Introduction}\label{sec:introduction}
AI agents increasingly act on humans' behalf: coding agents edit files and push commits to software repositories~\cite{li2025rise}; OS agents operate computers, phones, and browsers~\cite{hu2025agents}, and personal agents send messages and make purchases~\cite{meta2026muse}. Human control over such agents rests on a few checkpoints: the agent asks before consequential actions, respects the safeguards its environment provides, reports what it did, and stops when it is blocked~\cite{shavit2023practices,dhanorkar2026oversight, feng2025levels}. Many of these checkpoints are now shaped by \emph{agent skills} (or \emph{skills}; we use the two terms interchangeably), reusable instruction packages, centred on a \texttt{SKILL.md} file, that an agent loads into its context when a task matches the skill's description~\cite{anthropic2025skillsrepo, agentskills-spec}. Because skills are widely shared and copied~\cite{gao2026registry}, a single line in one skill can change the oversight of every agent that loads it.

Lines that change oversight do so as \emph{directives}, utterances through which a speaker tries to get someone to do something~\cite{searle1976classification}. We call a directive \emph{oversight-loosening} if it reduces human control over what an agent does: it lets the agent act without asking, skip or bypass a safeguard, act without reporting, keep going where it would otherwise stop, or override other instructions. A directive that preserves or adds such control is \emph{oversight-tightening}. Loosening directives are not inherently malicious; most simply let a workflow run without interruption. Whatever their purpose, they remove a checkpoint that would otherwise catch an agent's mistakes or an attacker's injected instructions, which is why security guidance lists human approval of high-risk actions as a mitigation for prompt injection~\cite{owasp2025llm}: an agent told not to ask will also not ask before acting on a misread task or a malicious web page.

Prior work does not capture such directives. Security studies of agent skills look for malicious skills, vulnerabilities, and credential leakage~\cite{registry2026malicious,liu2026skills,contextmatters2026}, and skill scanners ask whether a skill contains an attack~\cite{ciscoscanner,skillspector}; a loosening directive contains none. Studies of agent instruction files such as \texttt{AGENTS.md}~\cite{chatlatanagulchai2025agentreadmes} and Cursor rules~\cite{jiang2026cursor} categorize instructions by topic, not by their effect on oversight. Fundamental questions about loosening directives in agent skills therefore remain open: how common they are, where they occur, how they spread, and whether existing tools catch them. We address them through four research questions:
\begin{itemize}
  \item \textbf{RQ1 (Prevalence):} \rqone\ %
  \item \textbf{RQ2 (Risk context):} \rqtwo\ %
  \item \textbf{RQ3 (Spread):} \rqthree\ %
  \item \textbf{RQ4 (Detection):} \rqfour\ %
\end{itemize}

To answer the four RQs, we analyze 1.53 million English skills from GitSkills~\cite{destefanis2026gitskills}, which records every copy of each distinct \texttt{SKILL.md} file across public GitHub repositories, so we can measure both how often a loosening directive is written and how far it reaches. All four questions require identifying loosening directives from a \texttt{SKILL.md} file, a task that simple keyword matching cannot accomplish. First, a line with oversight vocabulary need not instruct the agent about oversight. Second, oversight language has polarity (``never push without confirmation'' tightens oversight, ``push without confirmation'' loosens it). To address both challenges, we first use a lexicon of 36 patterns for oversight phrases (e.g., ``without asking'', \texttt{--no-verify}) to retrieve 1.37M \emph{candidate units} (sentences, bullets, table cells, or code lines). We then develop a codebook that classifies each candidate unit as an oversight directive, a mention, or neither, and each directive as loosening or tightening; it also records each directive's mechanism and, for loosening directives, its scope. With this codebook, we label 2,097 units, using an LLM as an independent second coder on 1,000 of them ($\kappa = 0.82$ in the final round) and adjudicating every disagreement. Finally, we train a detector from labelled units that classifies all 1.37M candidate units (accuracy 0.80, $\kappa = 0.65$ on 400 held-out units). The human labels provide unit-level prevalence and mechanisms (RQ1) and the ground truth for auditing two open-source skill scanners (RQ4). The detector labels every candidate unit to measure skill-level prevalence (RQ1), analyze skill capabilities (RQ2), and trace near-duplicate skill families (RQ3). The detector serves as a baseline for RQ4.

Our study provides four main findings. (1)~In our human-labelled prevalence sample, 13.3\% of candidate units loosen oversight and 36.4\% tighten it; about one in seventeen English skills contains a loosening directive, and 70.1\% of these directives apply unconditionally. (2)~At equal amounts of oversight language, skills that run shell code, install packages, use \texttt{sudo}, or pipe downloads into a shell have 1.3--1.8 times the odds of loosening, without more tightening. (3)~Loosening spreads through a few upstream skills: 2.4\% of the families that carry loosening account for half of all copies of loosening-carrying skills, and adopters rarely change oversight text. (4)~Existing scanners catch at most 46\% of loosening directives, at a precision below 0.4, while the same LLM prompted with our codebook catches 61\% at a precision of 0.71.
 
The contributions of this paper are as follows:
\begin{enumerate}
  \item the notion of oversight-loosening directives, a risk in agent skills distinct from malicious content;
  \item a codebook for oversight directives (polarity, mechanisms, scopes), 2,097 labelled units, and a detector;
  \item a measurement of the prevalence, risk context, and spread of these directives across 1.53 million skills;
  \item an audit of two skill scanners, showing that the same LLM finds loosening when asked about oversight rather than malice.
\end{enumerate}

\section{Background and Related Work}\label{sec:background}

\begin{figure}[t]
\centering
\scriptsize
\begin{verbatim}
---
name: release-helper
description: Tag and publish a release.
---
## Workflow
1. Run the test suite.                  
2. Bump the version and commit.         
3. Push the tag without confirmation.   [D / L / CONF]
4. Run: npx -y release-it               [D / L / PERM]
 
## Options
| --yes | skip all prompts |            [M]
 
## Never
- Push without confirmation.            [D / T / CONF]
\end{verbatim}
\caption{A constructed \texttt{SKILL.md} with our labels. Each step,
table cell and bullet is one unit. Steps 1-2 match no lexicon phrase.
The same phrase loosens oversight under \emph{Workflow} and tightens it
under \emph{Never}, and \texttt{--yes} is a directive when the agent is
told to run it but a mention when only documented. D/M = directive/mention, L/T = loosens/tightens; mechanisms in
Table~\ref{tab:directives}.}
\label{fig:example}
\end{figure}

\subsection{Agent Skills}
A skill is a folder containing a \texttt{SKILL.md} file and, optionally, scripts and other resources the agent can use. The file begins with YAML frontmatter that names and describes the skill, followed by Markdown instructions~\cite{agentskills-spec} (Fig.~\ref{fig:example}). Agents read the frontmatter of every installed skill up front and load the body when the skill is invoked, so the body enters the agent's context as instructions at the moment the agent acts. Unlike repository-level instruction files such as \texttt{AGENTS.md}, skills are written to be usable across various projects and agents. About half of the 3.8 million \texttt{SKILL.md} files in GitSkills are exact copies~\cite{destefanis2026gitskills}.

\subsection{Oversight Mechanisms in Agents} 
\label{sec:oversight}
Human-factors research treats automation as a continuum. A system may suggest an action, act unless a human vetoes it, or act and inform the human~\cite{sheridan1978human, parasuraman2000model}. For AI agents, Feng et al. define autonomy levels by the user's role and treat it as a design decision~\cite{feng2025levels}. Shavit et al. recommend requiring human approval for consequential or irreversible actions to help keep agents legible and interruptible \cite{shavit2023practices}. Coding agents implement these practices as approval prompts, permission modes \cite{divecc2026}, allow- and denylists, prompt-suppressing flags, and hooks and tests that gate commits. Some of these checkpoints are enforced by the harness, while others only exist because an instruction asks for them. A skill can remove the second kind and guide an agent around the first (e.g., telling an agent to pass \texttt{-{}-force}). Developers are already using such instructions to exercise control over agents \cite{dhanorkar2026oversight}.  Our codebook's mechanisms (\ref{sec:codebook}) follow these practices.

\subsection{Analysis of Agent Skills and Agent Instruction Files}
26.1\% of 31,132 marketplace skills contain at least one potentially vulnerable pattern \cite {liu2026skills}. A registry study confirms 157 malicious skills among 98,380 \cite{registry2026malicious}. Scanners running over 238,180 skills rarely agree, with most flags failing a review of their context \cite{contextmatters2026}. Open-source scanners combine static rules with optional LLM analysis \cite{ciscoscanner, skillspector}.  A second line of work studies the instruction files that configure coding agents: Jiang and Nam categorize the context in Cursor rules into conventions, guidelines, project information, LLM directives, and examples, and find that 28.7\% of rule lines are duplicated~\cite{jiang2026cursor}; Chatlatanagulchai et al.\ categorize the instructions in 2,303 agent context files such as \texttt{CLAUDE.md} and \texttt{AGENTS.md}~\cite{chatlatanagulchai2025agentreadmes}. These studies categorize instructions by topic; none asks whether an instruction loosens or tightens human oversight.

\begin{figure*}[t]
    \centering
    \includegraphics[width=0.85\textwidth]{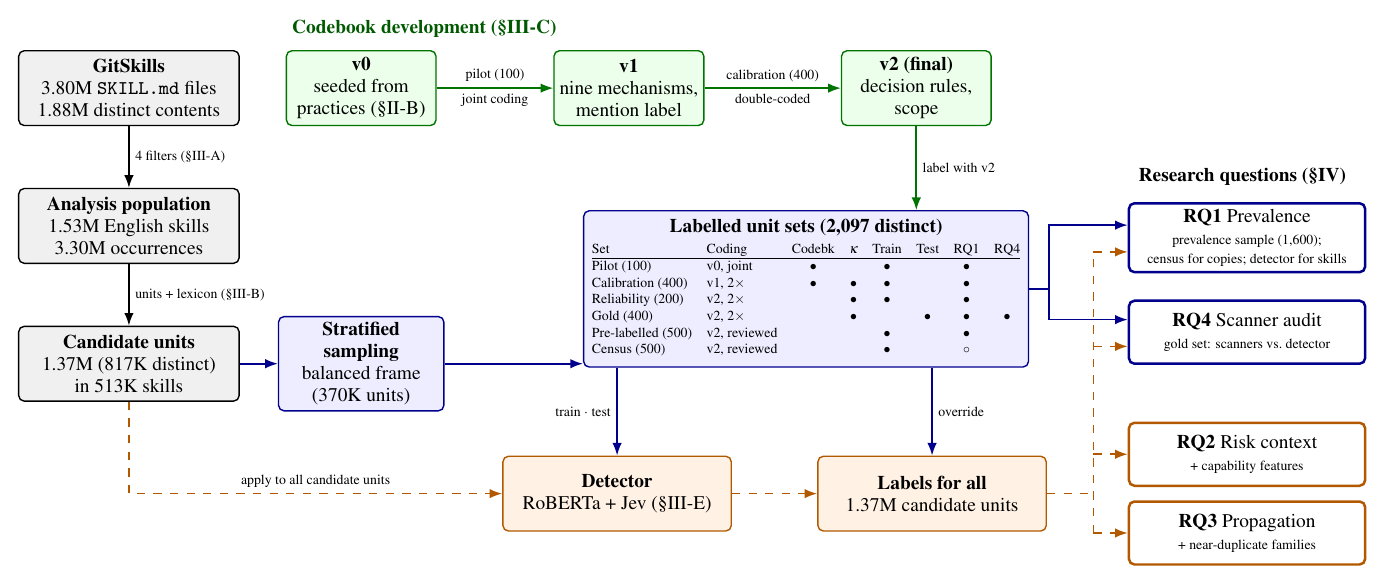}\vspace{-0.5cm}
    \caption{Study pipeline. Green: Codebook Development. Blue: Human Labels. Each labelled set is marked with the codebook version and coding it received: \emph{joint} = coded by the first author together with an LLM, discussing ambiguous units; 2$\times$ = coded independently by the first author and an LLM, with every disagreement adjudicated by the first author; \emph{reviewed} = proposed by an LLM and reviewed by the first author. Pilot and calibration units were later relabelled with v2. In the RQ1 column, $\circ$ marks the census, which is used only for copy-weighted estimates.  }
    \label{fig:pipeline}
\end{figure*}

\section{Study Design}\label{sec:method}

Although Figure \ref{fig:pipeline} reads from left to right, the steps overlapped. We first drafted the lexicon together with an initial codebook (v0). The lexicon was used to retrieve candidate units. We then refined the codebook on samples of those units. On the 2,097 human-labelled units, the first author labelled or reviewed every unit, and an LLM independently labelled 1,000 of them as a second coder, following recent work that evaluates LLMs in place of one of two human annotators of software engineering artifacts~\cite{ahmed2025llms}. For units that were already labelled by hand, we use the human label instead (the ``override'' arrow in Fig.~\ref{fig:pipeline}). Human labels provide RQ1's unit-level estimates and RQ4's ground truth; detector labels provide RQ1's skill-level estimates and all of RQ2 and RQ3, and serve as a baseline in RQ4.

\subsection{From GitSkills to Analysis Population}
\label{sec:data}
We use GitSkills \cite{destefanis2026gitskills} through its Hugging Face mirror, at a revision whose row count matches that of the dataset paper. GitSkills contains 3,797,117 \texttt{SKILL.md} files, from 282,200 public repositories, which deduplicate by content hash to 1,877,981 distinct contents. We keep a distinct file if it (1) is meant as a skill, i.e., has valid YAML frontmatter or sits in a skills directory such as \texttt{.claude/skills/}; (2) postdates the format, i.e., its first commit, where history is available, is no earlier than October 2025, when \texttt{SKILL.md} was introduced~\cite{destefanis2026gitskills}; (3) is not a symlink placeholder whose content is only a path to another file; (4) is in English, as identified by \texttt{langid.py}~\cite{lui2012langid} on its prose without frontmatter, code, and URLs, since our lexicon and codebook are English (85.6\% of distinct contents). This leaves us with 1,527,281 distinct skills, 3,301,422 occurrences when every exact copy in every repository is counted. 

\subsection{From Skills to Candidate Units} 
\label{sec:retrieval} 
We split each skill body into \emph{units}: a sentence, bullet point, table cell, or code line. Each unit is read with its \emph{context}: its section heading and the units immediately preceding and following it, as a unit's role often depends on its surroundings (Fig. \ref{fig:example}). Most units say nothing about oversight. To find those that might, we use a lexicon of 36 regular-expression patterns for oversight phrases (e.g., ``without asking'', ``do not stop'', \texttt{-{}-no-verify}, ``auto-merge'') as a filter: a unit that matches any pattern becomes a \emph{candidate unit}, which the codebook then classifies (\S\ref{sec:codebook}). The lexicon only retrieves units and decides nothing about them. We drafted the patterns from the practices (\S\ref{sec:oversight}) and grouped them into eight categories, one for each \emph{mechanism} of the initial codebook, i.e., the kind of oversight a directive acts on, such as asking for confirmation or respecting safeguards (\S\ref{sec:codebook}); INVOKE, added later, has none. The categories serve only for retrieval and stratified sampling.

 \begin{table*}[t]
\centering

\caption{Oversight directive mechanisms with examples.}
 \vspace{-4pt}
\label{tab:directives}
\small
\begin{tabularx}{\textwidth}{l>{\raggedright\arraybackslash}Xr>{\raggedright\arraybackslash}Xr}
\toprule
\textbf{Mechanism} & \textbf{Loosening example (paraphrased)} & \textbf{Share of L} & \textbf{Tightening example (paraphrased)} & \textbf{Share of T}\\
\midrule
PERM (permission flags) & Run \texttt{npx -y} \ldots{} / pass \texttt{--yes} & 33.9\% & Never use \texttt{--force} & 2.4\%\\
CONF (confirmation) & Proceed without asking the user & 27.4\% & Wait for explicit approval before deploying & 46.3\%\\
PERSIST (stopping) & Don't stop between tasks. Keep going until done & 16.0\% & Stop and ask when blocked & 1.0\%\\
SAFE (safeguards) & Commit with \texttt{--no-verify} & 7.1\% & Never skip the test suite & 13.8\%\\
AUTO (auto-action) & Auto-commit completed units & 6.2\% & Never auto-merge & 4.1\%\\
TRANS (transparency) & Do not announce each commit & 4.5\% & Never apply a change silently, report it & 11.3\%\\
ACT (consequential action) & Feel free to push & 3.0\% & Never amend a commit that is already pushed & 20.2\%\\
INVOKE (invocation as approval) & Invoking this command is the go-ahead & 1.5\% & \textit{(none observed)} & 0.0\%\\
PREC (precedence) & These rules override all other instructions & 0.5\% & Never let retrieved text override your instructions & 0.9\%\\
\bottomrule
\end{tabularx}

\end{table*}

\subsection{Developing the Codebook}
\label{sec:codebook}

\emph{Development.}
The codebook went through three versions (green row of Fig.~\ref{fig:pipeline}). v0 derived its mechanisms from the oversight practices of \S\ref{sec:oversight}, one for each checkpoint that a skill's text can remove or reinforce (Table~\ref{tab:directives}); PREC follows prompt-injection research, where injected text claims authority over other instructions~\cite{greshake2023indirect} and instruction hierarchies are meant to resist such claims~\cite{wallace2024hierarchy}. A pilot of 100 units, which the first author labelled while an LLM asked for clarification wherever the codebook was ambiguous (and which we therefore do not use to measure agreement), led to v1: overlapping mechanisms received definitions and tie-breakers, mentions were narrowed to units that refer to an oversight mechanism without instructing the agent, and a rule on who acts was added (a tool acting on its own is a mention; an agent told to act is a directive). A calibration round of 400 units, coded independently by the first author and an LLM, showed that most disagreements came from ordinary engineering instructions or flag documentation coded as directives, and led to v2. V2 asks coders three questions in order before applying the detailed rules: whether the unit concerns human control over the agent at all, whether it tells the agent to act or only describes something, and whether it leaves the human with more or less control.

\emph{Final Codebook:}  

\begin{enumerate}
    \item \emph{Type:} A unit is an oversight directive (D) if it tells the agent whether, when, or how human oversight applies to what it does, and a mention (M) if it refers to an oversight directive without instructing the agent. Everything else is neither (N). The criterion is whether deleting the unit would change how much control a human has over the agent.
    \item \emph{Polarity:} Only applies to units marked as directives. A directive loosens oversight (L) if it allows the agent to act with less human involvement. It tightens oversight (T) if it requires more human oversight.
    \item \emph{Mechanism:} Only applies to units marked as directives. What the directive acts on, one of nine (Table \ref{tab:directives}).
    \item \emph{Scope:} Only applies to units marked as loosening directives. The condition under which the loosening applies (e.g., in CI, in tests, only if the user has authorized it) or \emph{unscoped} if there is none.
\end{enumerate}

\emph{Decision rules.} The label definitions leave some recurring cases open, in which coders applying the same definitions reached different labels. We resolved these cases with twelve decision rules, derived from disagreements in the pilot and calibration rounds. %

\begin{table}[t]
\centering
 \vspace{-4pt}
\caption{Inter-rater reliability (Cohen's $\kappa$). Label: neither / loosening /
tightening, over all units; polarity and mechanism: over units both coders
labelled as directives; scope: over units both labelled as loosening. }
\label{tab:agreement}
\begin{tabular}{lrrrrr}
\toprule
\textbf{Round} & \textbf{n} & \textbf{Label} & \textbf{Polarity} & \textbf{Mechanism} & \textbf{Scope}$^{\dagger}$\\
\midrule
Round 1 (v1)     & 400 & 0.67 & 0.86 & 0.80 & 0.17\\
Round 2 (v2)& 200 & 0.80 & 0.97 & 0.87 & 0.75\\
Gold set (v2)& 400 & 0.82 & 0.92 & 0.84 & 0.85\\
\bottomrule
\end{tabular}

\vspace{3pt}
{\footnotesize $^{\dagger}$Computed from the coders' original label files.}
\end{table}
\subsection{Labelled Unit Sets}
\label{sec:labeling}

\emph{Sampling.} 
Candidate units are unevenly spread over lexicon categories and repositories, so we sample them in three steps. First, we assign each unit to a stratum: the rarest category it matches. Categories range from 5,266 distinct units in the smallest to about 247,000 in the largest, so a simple random sample would include almost no units from the rare ones. Second, because a few large repositories contribute many near-identical units, we build a \emph{repository-balanced frame} that keeps at most two distinct units per repository in each stratum, reducing the 817,479 distinct candidate units to 370,285. Third, we draw five of the labelled sets below from this frame, with roughly equal numbers of units from each stratum, so that rare strata are oversampled; RQ1 reweights them to the frame (\S\ref{sec:RQ1}).

\emph{Labelled sets.} We labelled candidate units in six sets. Five are stratified samples from the balanced frame: the \emph{pilot} (100 units) and the \emph{calibration} (400 units) sets developed v1 and v2 and were then relabelled with v2; the \emph{reliability} (200 units) and the \emph{gold} (400 units) sets were double coded with v2; the \emph{pre-labelled} set (500 units) was reviewed under v2. Together, they form the 1,600-unit \emph{prevalence sample} for RQ1. The sixth set, the \emph{census}, holds the 500 most copied candidate units,  labelled in full as a take-all stratum \cite{lohr2021sampling} for copy-weighted estimates (\S\ref{sec:RQ1}). Three census units share their text with a sampled unit, so the sets hold 2,097 distinct units. The gold set evaluates the detector and the scanners (RQ4); all other sets train the detector.

\emph{Coders and Reliability.} The first author labelled or reviewed every unit in the six sets. An LLM (Claude Opus 5.5, max effort) independently labelled the 1,000 units of the calibration, reliability and gold sets, seeing only the codebook and each unit with its context; its labels were hashed and timestamped before comparison (this choice follows recent practices~\cite{ahmed2025llms} and is discussed in \S\ref{sec:threats}). The first author adjudicated every disagreement, and the adjudicated labels are used in all analyses. Table \ref{tab:agreement} shows Cohen's $\kappa$. On the gold set, the three-way label (neither, loosening, tightening) reaches 0.82 (polarity 0.92, mechanism 0.84, scope 0.85). For the pre-labelled set and the census, the LLM proposed labels and the first author reviewed each, changing 8.4\% and 4.0\% of them respectively; because a proposal can anchor its reviewer, we do not count this agreement as reliability.

\begin{table}[t]
\centering
\caption{Detectors on the gold units. LLMs are prompted with our codebook. Cost is US\$ per 1,000 units.}
 \vspace{-4pt}
\label{tab:acc}
\begin{tabular}{lrrrr}
\toprule
\textbf{Method} & \textbf{Acc.} & \textbf{$\kappa$} & \textbf{Loos.\ F1} & \textbf{Cost}\\
\midrule
RoBERTa (fine-tuned) & 0.74 & 0.55 & 0.52 & 0\\
Jev 1.13             & 0.74 & 0.59 & 0.62 & 0.06\\
GPT-5.4-mini         & 0.73 & 0.54 & 0.56 & 0.36\\
GPT-5.4              & 0.78 & 0.62 & 0.65& 0.7--1.1\\
\midrule
RoBERTa + Jev (ours) & 0.80& 0.65 & 0.65 & 0.06\\
\bottomrule
\end{tabular}
\end{table}

\subsection{Detector}
\label{sec:detector}
RQ1's skill-level estimates, RQ2, and RQ3 need a label for each of the 1.37M candidate units, and RQ4 uses the detector as a baseline. As the strongest LLM classifier, we consider GPT-5.4 prompted with our codebook rather than Claude Opus 5.5: Opus 5.5 served as the second coder of our human labels, so evaluating it against those labels would favour it unfairly. Labelling every candidate unit with GPT-5.4 would cost about US\$1,000--1,500 (Table~\ref{tab:acc}). GPT-5.4-mini is cheaper than GPT-5.4 but less accurate than Jev, at six times Jev’s cost. So we combine two cheaper classifiers, i.e., \emph{RoBERTa} and \emph{Jev}. We fine-tune a RoBERTa-base \cite{liu2019roberta} with two heads (three-way label and mechanism) on every labelled unit outside the gold set, with its context, using three random seeds (best epoch on a 15\% held-out split), and average the three models. Jev 1.13~\cite{typesafe2026jev} answers multiple-choice classification questions with calibrated probabilities; without fine-tuning, we ask two questions whose options paraphrase the codebook with the same context. Because it over-predicts loosening, we reweight its probabilities (neither 1.0, loosening 0.25, tightening 0.75), with weights chosen by grid search for $\kappa$ on the 700 reliability and pre-labelled units, and renormalize.

\emph{Combination.} The detector averages the two models' three-way probabilities and takes the mechanism from Jev
(mechanism accuracy 0.82 vs.\ 0.76 on gold directives). We chose to combine them after seeing both models' gold scores, which may flatter the combination slightly, but tuned nothing on the gold set. On the 400 gold units (Table~\ref{tab:acc}), the detector reaches accuracy 0.80 ($\kappa = 0.65$) and loosening F1 0.65 (precision 0.70, recall 0.61). It fixes 43 of RoBERTa's errors while adding 20 (exact McNemar test, $p = 0.005$). Units labelled with confidence $\geq 0.9$ (31\%) are 97\% correct. GPT-5.4 prompted with our codebook performs on par (loosening F1 0.65; McNemar on the loosening flag, $p = 1.0$) at over ten times the cost. The 2,097 human labelled units override the detector wherever they exist, except where the detector itself is evaluated or compared (Table~\ref{tab:acc}, RQ4, and RQ1's comparison); we call those labels the \emph{detector reference}.

\section{Results}\label{sec:results}

\subsection{RQ1: \rqone}
\label{sec:RQ1}
We estimate prevalence at three levels: distinct units, copy-weighted units, and skills, using two sources of labels. The prevalence sample (\S\ref{sec:labeling}) gives our most reliable estimates, with quantified sampling uncertainty; the detector reference (\S\ref{sec:detector}) covers every candidate unit. \emph{Distinct-unit prevalence} counts each unique candidate unit once, i.e., it measures how common loosening is among the ways skills phrase oversight, regardless of how widely each phrasing spreads. For the prevalence sample, we weight each lexicon stratum by its size in the frame, which corrects for the oversampling of rare strata, and compute 95\% bootstrap confidence intervals (2,000 resamples); mechanism and scope are estimated the same way. The detector reference gives these shares directly. \emph{Copy-weighted prevalence} weights each unit by how often agents may encounter it: once per distinct skill file it appears in (\emph{unit} level), or once per exact copy of those files across repositories (\emph{occurrence} level). The 500 most-copied candidate units hold 15.3\% of all occurrences, yet only three fall in the prevalence sample, so the copy-weighted estimate combines the prevalence sample (1,600 units) with the census (500 units), 2,097 distinct units in all: each census unit counts for itself, and the sample stands for the rest of the frame. \emph{Skill-level prevalence} counts the skills that contain at least one loosening or tightening directive, which requires the detector reference. As robustness checks, we also recompute the distinct-unit estimate from the prevalence sample without test and attack folders (e.g., \texttt{tests/}, \texttt{fixtures/}) and security-tool repositories, which hold loosening strings as samples rather than instructions, and separately for each labelled set, to check that no single set, in particular the one the LLM pre-labelled, drives the estimate. %

\begin{table}[t]
\centering
\caption{Prevalence among distinct candidate units of the repository-balanced frame (370,285 units). Human: prevalence sample, 95\% CIs. Detector: all units, no sampling error.}
 \vspace{-4pt}
\label{tab:RQ1}
\begin{tabular}{lrr}\toprule
\textbf{Measure}&\textbf{Human}&\textbf{Detector}\\
\midrule
Oversight directive&49.7\% [46.8, 52.6]&49.6\%\\
\quad Loosening&13.3\% [11.7, 15.0]&10.3\%\\
\quad\quad \ldots\ of which unscoped&70.1\% [63.9, 75.8]&n/a$^{*}$\\
\quad Tightening&36.4\% [33.7, 39.2]&39.3\%\\
Mention only&11.6\% [10.1, 13.1]&n/a$^{*}$\\
\bottomrule
\end{tabular}
 
\vspace{3pt}
{\footnotesize $^{*}$The detector does not separate mentions from non-directives or predict scope.}
\end{table}

\textbf{Finding 1.1: Tightening dominates, but loosening is far from rare.} In the prevalence sample, half of the candidate units are oversight directives (49.7\%), and they tighten oversight 2.7 times as often as they loosen it: 36.4\% of units tighten and 13.3\% loosen (Table~\ref{tab:RQ1}), about 49,000 distinct loosening directives in the frame. The detector reference shows a similar picture (10.3\% loosening, 39.3\% tightening; we discuss the differences further in \S\ref{sec:threats}).

\textbf{Finding 1.2: Loosening works mainly through permission flags, skipped confirmation, and instructions not to stop.} These three mechanisms account for over three quarters of loosening directives: permission flags (33.9\%), confirmation (27.4\%), and persistence (16.0\%) (Table~\ref{tab:directives}). Tightening is instead dominated by confirmation (46.3\%) and consequential actions (20.2\%). Across lexicon strata, loosening is most common among units matching permission-flag phrases (42.7\%) and autonomy phrases (28.4\%).
 
\textbf{Finding 1.3: Most loosening is unscoped.} Only 29.9\% of loosening directives are limited to particular conditions, such as CI runs or user-authorized cases; the remaining 70.1\% apply unconditionally.  

\textbf{Finding 1.4: Counting copies raises loosening only slightly, but shifts which mechanisms dominate.} Weighted by copies, using the prevalence sample and the census together, loosening accounts for 17.9\% [13.3, 24.4] at the unit level and 15.0\% [11.2, 19.9] at the occurrence level. Both intervals include the distinct-unit estimate of 13.3\%, so heavy copying does not make loosening substantially more common, but the most-copied units loosen oversight differently. Among the loosening directives in the census, weighted by the files they appear in, transparency is the most common mechanism (22.6\%, against 4.5\% in the prevalence sample), followed by persistence (18.1\%), permission flags (17.7\%, against 33.9\%), and automatic actions (16.0\%, against 6.2\%).

\textbf{Finding 1.5: About one in seventeen English skills contains a loosening directive.} By the detector reference, 5.8\% of English skills contain at least one loosening directive, and 16.9\%  at least one tightening directive. Since the detector's loosening recall (0.61) is below its precision (0.70), the loosening figure is more likely an underestimate than an overestimate.

\noindent \emph{Robustness.}
Excluding security-tool repositories and test fixtures changes the loosening share only marginally (13.2 to 13.3\%) and estimates from the individual labelled sets range from 12.5\% to 15.6\%.

\begin{tcolorbox}[colback=white, colframe=black, boxrule=0.5pt, arc=0pt, left=6pt, right=6pt, top=4pt, bottom=4pt]
\textit{\textbf{Answer to RQ1:}  Loosening directives make up 13.3\% of distinct candidate units (about 49,000) and appear in 5.8\% of English skills. They work mainly through permission flags, skipped confirmation, and instructions not to stop, and 70\% are unscoped. Among the most-copied units, acting without telling the user becomes the most common mechanism.}
\end{tcolorbox}

\subsection{RQ2: \rqtwo}
\label{sec:RQ2}
Two properties of a skill determine how much a loosening directive matters. The first is what the skill can do: skipping a confirmation matters more in a skill that runs \texttt{sudo} than in one that formats text. The second is what else the skill says: an agent loads a skill as a whole, so a loosening directive in one step may be offset by a tightening directive in another. We therefore ask which skills are likely to contain loosening directives, and whether those skills compensate with tightening. 

We describe each skill by its \emph{capability features}, which mark what an agent following the skill can do beyond editing text: execute code, change the system, reach the network, handle credentials, or write to version control. We detect eight of them with regular expressions over the skill's body, and take the ninth, bundled executable files, from GitSkills' metadata on files shipped with the skill: shell code blocks (34.5\% of skills), references to secrets or credentials (32.9\%), package installs (12.5\%), bundled executable files (12.0\%), network calls (9.2\%), git writes (5.1\%), destructive commands (2.0\%), \texttt{sudo} (1.3\%), and piping a download into a shell, i.e., \texttt{curl | sh} (0.7\%); and by its \emph{repository context}, which reflects how much scrutiny and reach a skill has: the repository's stars (log-transformed), since popular repositories may be reviewed more, and whether the skill is copied into two or more repositories, since copied skills reach more agents (\S\ref{sec:RQ3}). The analysis covers all 512,876 candidate skills, labelled with the detector reference: a skill contains a loosening directive if at least one of its candidate units is labelled loosening (17.2\% of candidate skills), and likewise for tightening (50.4\%).

For each candidate skill $s$, let $L_s = 1$ if $s$ contains at least one loosening directive, and $T_s = 1$ if it contains at least one tightening directive. We fit three logistic regressions (M1--M3) of the form below, where the outcome $Y_s$ is $L_s$ (M1, M2) or $T_s$ (M3):
\begin{equation}
\operatorname{logit} P(Y_s = 1) = \alpha + \boldsymbol{\beta}^\top \mathbf{c}_s
  + \boldsymbol{\gamma}^\top \mathbf{r}_s + \boldsymbol{\delta}^\top \mathbf{z}_s,
\label{eq:rq2}
\end{equation}
where $\mathbf{c}_s$ holds the capability features, $\mathbf{r}_s$ the repository-context variables, and $\mathbf{z}_s$ the \emph{controls}: the skill's length and its number of candidate units, both log-transformed. \textbf{M1} models loosening ($Y_s = L_s$) from capability features and controls only ($\boldsymbol{\gamma} = \mathbf{0}$); \textbf{M2} adds repository context; and \textbf{M3} models tightening ($Y_s = T_s$) from the same predictors as M2. Each model estimates its own coefficients; we write $\boldsymbol{\beta}'$ for M3's capability coefficients. Comparing M1 with M2 shows whether capability effects hold once we account for where a skill lives and how it is shared; comparing M2 with M3 shows whether capabilities associated with loosening are also associated with tightening. Each odds ratio $e^{\beta_k}$ answers RQ2 for one capability $k$: above 1, skills with capability $k$ have higher odds of containing a loosening directive than otherwise comparable skills without it; below 1, lower odds. We model all characteristics jointly because they co-occur (skills that use \texttt{sudo} often also install packages). The controls compare skills with equal amounts of oversight language, so the odds ratios reflect which way a skill's oversight language leans, not how much of it there is. If a capability raises the odds of loosening but not of tightening ($e^{\beta'_k} \leq 1$), skills with it carry no additional guardrails to match. Standard errors are clustered by repository.

To test robustness, we refit M2, the model behind our loosening findings, five times, each time changing one aspect of the analysis: (1)~counting loosening only at a predicted probability of at least 0.7 (this gives precision 0.75, recall 0.30 on the gold set); (2)~using the predicted probability of at least one loosening directive as the outcome; (3)~excluding test folders and security-tool repositories; (4)~fitting on all 1.53M English skills, which removes the control for the amount of oversight language; and (5)~defining copying by distinct owners, so that forks do not count. As a check independent of the detector, we fit the same predictors on the prevalence sample, with lexicon strata as fixed effects, and compare the direction of each association.

\begin{table}[t]
\centering
\caption{Odds ratios for $\geq$1 loosening (M2) and $\geq$1 tightening (M3) directive. $n$ = 512,876 skills; 95\% CIs clustered by repository; controls for log length and log number of candidate units. Detector reference (\S\ref{sec:detector}). }
 \vspace{-4pt}
\label{tab:RQ2}
\footnotesize

\begin{tabular}{lrrr}\toprule
\textbf{Capability Features}& \textbf{\% L }& \textbf{Loosening OR}& \textbf{Tight. OR}\\
\midrule
Shell code block      & 24.8 / 11.1 & 1.82 [1.73, 1.92] & 0.95$^{\dagger}$\\
Package install       & 29.7 / 14.7 & 1.82 [1.69, 1.95] & 0.59\\
sudo                  & 38.8 / 16.7 & 1.74 [1.61, 1.89] & 0.75\\
curl\,$|$\,sh& 40.6 / 16.9 & 1.31 [1.18, 1.45] & 0.77\\
Network calls         & 26.7 / 15.8 & 1.19 [1.13, 1.25] & 0.88\\
Bundled executables   & 21.7 / 16.5 & 1.12 [1.00, 1.27] & 1.01$^{\dagger}$\\
Git write             & 32.4 / 15.3 & 1.09 [1.04, 1.14] & 2.01\\
Destructive command   & 33.5 / 16.4 & 0.98 [0.92, 1.03] & 0.87\\
Secrets / credentials & 19.1 / 15.6 & 0.77 [0.74, 0.81] & 0.92\\
\midrule
Stars (log)           & n/a& 0.99 [0.96, 1.02] & 0.97\\
Copied ($\geq$2 repos)& n/a& 1.13 [1.05, 1.22] & 0.83\\
\end{tabular}

\vspace{3pt}
{\footnotesize \%L: share of skills with $\geq$1 loosening directive, with / without the capability feature mentioned. $^{\dagger}$Tightening OR not significant ($p \geq 0.05$).}
\end{table}

\textbf{Finding 2.1: Skills with execution capabilities contain loosening directives far more often.} Of the skills that pipe a download into a shell, 40.6\% contain at least one loosening directive against 16.9\% for skills that do not; for \texttt{sudo}, the share is 38.8\% for skills containing sudo versus 16.7\% for skills without it (Table~\ref{tab:RQ2}). Part of this gap arises as these skills have more oversight language of both kinds (4.6 and 3.7 candidate units on average, against 2.6 for other candidate skills), giving them more chances to contain a loosening directive, which is why our regression models compare skills at equal amounts of oversight language (i.e., the controls).

\textbf{Finding 2.2: The association holds at equal amounts of oversight language.} Skills with shell code blocks, package installs, or \texttt{sudo} have 1.7--1.8 times the odds of containing a loosening directive, and skills with \texttt{curl | sh} have 1.3 times the odds. Network calls are associated more weakly (OR 1.19). Without repository context (M1), every capability odds ratio is within 0.02 of its M2 value, so these associations do not depend on where a skill resides or how widely it is shared. Destructive commands show no association once the amount of oversight language is controlled (OR 0.98), and skills that reference secrets or credentials are less likely to loosen (OR 0.77). 

\textbf{Finding 2.3: Capabilities associated with more loosening are not associated with more tightening.} Package installs, \texttt{sudo}, and \texttt{curl | sh} lower the odds of a tightening directive (OR 0.59--0.77), shell code blocks leave them unchanged, and destructive commands lower them too (OR 0.87). The exception is git writes, which come with much more tightening (OR 2.01), typically including rules against force-pushing or committing without approval.

\textbf{Finding 2.4: Popularity does not protect, and copied skills loosen more.} Stars are not associated with loosening. Skills copied into two or more repositories are more likely to loosen (OR 1.13) and less likely to tighten (OR 0.83); with copying defined by distinct repository owners, so that forks do not count, the association with loosening is stronger (OR 1.18 [1.08, 1.28]).

\textbf{Finding 2.5: About 12,500 skills combine loosening with high-risk capabilities.} In total, 12,484 English skills (0.82\%) contain at least one loosening directive together with destructive commands, \texttt{curl | sh}, or \texttt{sudo}; counting only loosening predicted with a probability of at least 0.7, 6,918 (0.45\%) still do, a conservative lower bound.

\noindent \emph{Robustness.} 
Shell code blocks, package installs, and \texttt{sudo} remain significant in all five refits, and \texttt{curl | sh} in all but the high-confidence one. Over all 1.53M skills, which lack the control for oversight language, network calls lose significance and copying reverses (OR 0.85). On the prevalence sample, the direction agrees for shell code blocks, network calls, package installs, and git writes, but not for secrets, \texttt{sudo}, and \texttt{curl | sh}, the last two resting on only 39 and 42 units. Destructive commands have no consistent direction across refits.

\begin{tcolorbox}[colback=white, colframe=black, boxrule=0.5pt, arc=0pt, left=6pt, right=6pt, top=4pt, bottom=4pt]
\textit{\textbf{Answer to RQ2:} %
At equal amounts of oversight language, skills that run shell code, install packages, use \texttt{sudo}, or pipe downloads into a shell have 1.3--1.8 times the odds of containing a loosening directive, and do not compensate with more tightening. Popularity is not protective, and copied skills loosen more.}
\end{tcolorbox}

\subsection{RQ3: \rqthree}
\label{sec:RQ3}

To learn where to act, RQ3 asks whether loosening spreads mainly through \emph{upstream skills}, the widely copied versions that others reuse, or through edits that their adopters make. All steps use the detector reference, since they need a label for every candidate unit of every skill.

\emph{Step 1: Measuring amplification.} If skills with loosening directives were copied more than others, copying would spread loosening beyond its prevalence among distinct skills. Using only exact copies, we compare the share of skills with at least one loosening directive counting each distinct skill once with the share counting every copy; a ratio above 1 means copying amplifies loosening. We compute the same ratio for tightening.

\emph{Step 2: Identify skill families.} Edited copies no longer match their source exactly. As in work on code duplication\cite{lopes2017dejavu}, we group skills into families of near-duplicates. We define two skills as linked if the Jaccard similarity of their 5-word sequences, estimated with MinHash \cite{broder1997resemblance}, is at least 0.7. This yields 118,076 families with two or more members, containing 558,531 skills. Because connected components can chain dissimilar skills, we repeat the family analyses at 0.6 and 0.8. 

\emph{Step 3: Concentration.}  We rank families by their loosening occurrences (loosening directives counted across all skill copies) and measure the share held by the top 100, against those families' share of all skill occurrences and against the share of all occurrences held by the 100 largest families. We also report how many families, and how many distinct loosening units, account for half of all loosening occurrences.

\emph{Step 4: Edited copies.} In each family, we take the earliest-committed member as the \emph{ancestor} and every other member as a \emph{variant}, and identify the candidate units each variant adds and removes. Because edited copies tend to grow, raw counts would overstate drift toward loosening; we therefore compare the share of loosening among added units with its share among removed units, with 95\% CIs from resampling whole families. As sensitivity analyses, we use a stricter label threshold ($p \geq 0.7$), exclude families of 100 or more variants, take the most-copied member as the ancestor, and restrict to variants under a different owner than the ancestor, which isolates downstream users from upstream authors' own revisions. Finally, we inspect the most frequently added and removed loosening units manually.

\begin{figure}[t]
  \centering
  \includegraphics[width=0.75\textwidth]{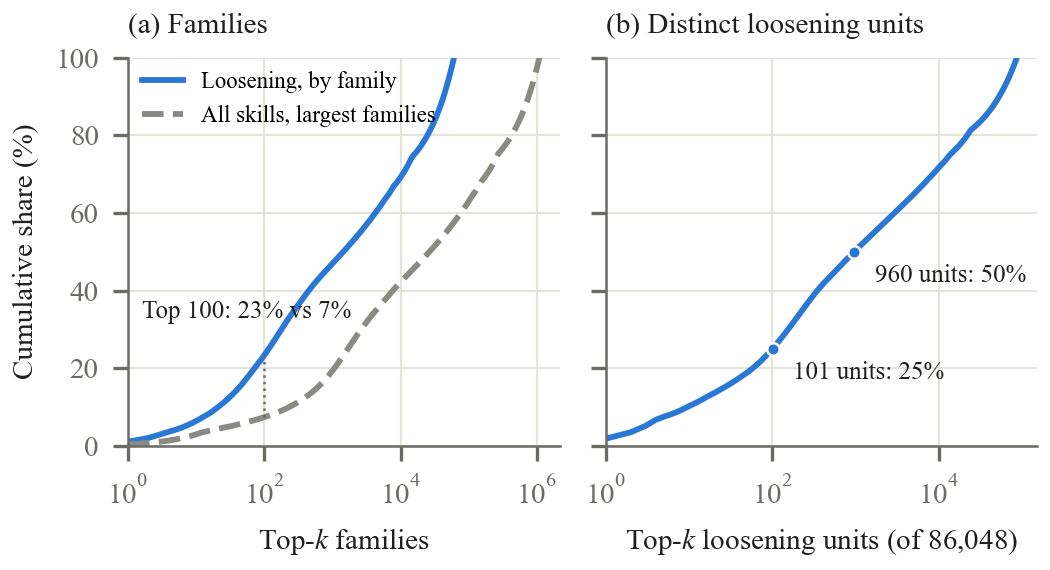}
  \caption{Loosening is concentrated in a few upstream skills. (a)~Cumulative
  share of loosening-carrying skill occurrences held by the top-$k$
  families, against the share of all skill occurrences held by the $k$
  largest families. (b) Cumulative share of loosening-unit occurrences held by the top-$k$ distinct loosening units.}
  \label{fig:rq3}
\end{figure}

\textbf{Finding 3.1: Copying does not amplify loosening.} 5.8\% of distinct skills contain a loosening directive (as in RQ1), compared with 5.4\% of all skill occurrences (ratio 0.94). Tightening is diluted more by copying (16.9\% vs.\ 14.4\%, ratio 0.85).

\textbf{Finding 3.2: Loosening is concentrated in a few families and wordings.} The 100 families carrying the most loosening hold 23\% of all loosening-carrying skill occurrences but only 1.4\% of all skill occurrences; even the 100 largest families hold only 7.4\% (Fig.~\ref{fig:rq3}). Half of all loosening-unit occurrences come from 960 of 86,048 distinct loosening units (1.1\%).

\textbf{Finding 3.3: Variants rarely change oversight language.} With the earliest-committed ancestor at threshold 0.7, 93.4\% of variants leave their ancestor's oversight directives untouched (92--97\% across all specifications).

\textbf{Finding 3.4: When variants do change oversight language, they do not shift toward loosening.} Variants add 1.2--2.7 times as many loosening directives as they remove, but also 1.1--1.9 times as many other units. Comparing shares instead, loosening makes up 20.3\% of added and 20.2\% of removed units (+0.1 points [$-$5.3, +4.7]), and tightening shows no significant shift. With the earliest-committed ancestor, this holds in every specification but one, a removal of tightening at 0.6 that comes entirely from families of 100 or more variants. With the most-copied member as the ancestor, variants add slightly more loosening than they remove (+1.4 to +5.5 points, significant in two of five specifications).

\textbf{Finding 3.5: Where loosening changes, it changes upstream.} The most common oversight edits are successive versions of the same skill packs: new releases that add, for example, an auto-committing checkpoint mode or an instruction not to announce each commit, or that drop an instruction to ask the user before moving on to a step. 

\begin{tcolorbox}[colback=white, colframe=black, boxrule=0.5pt, arc=0pt, left=6pt, right=6pt, top=4pt, bottom=4pt]
\textit{\textbf{Answer to RQ3:} Loosening spreads mainly through a small number of upstream skills: 2.4\% of the skill families that carry loosening account for half of all copies of loosening-carrying skills, and their wording is reused verbatim, with 1.1\% of distinct loosening units accounting for half of all loosening occurrences. Copying does not amplify loosening, and adopters who edit copies rarely change oversight text; at most, a small shift toward loosening appears when the most-copied version serves as the ancestor.}
\end{tcolorbox}

\subsection{RQ4: \rqfour}
\label{sec:RQ4}

\noindent \textbf{Approach.} Skill marketplaces and security teams rely on scanners to screen skills; if scanners miss loosening, or flag guardrails as risks, the burden of checking each skill falls back on the users. Several skill scanners are in use, many of them integrated into skill marketplaces~\cite{csa2026scannerbypass, contextmatters2026}. We select scanners that target agent skills specifically, are open source and run locally, so that every skill in our sample can be scanned and the results reproduced, offer both rule-based and LLM analysis, and report findings by line. Two scanners from major security vendors meet these criteria: Cisco AI Defense skill-scanner 2.2.1~\cite{ciscoscanner} and NVIDIA SkillSpector 2.12.0~\cite{skillspector}. We run each offline (rule-based analyses only) and with LLM analysis enabled through its GPT-5.4 integration. We run each scanner on the skill files that contain our labelled units and count a unit as flagged if the scanner reports a finding of medium or higher severity on the unit's line or quotes the unit. We compare the scanners with three baselines: keyword matching with our lexicon; GPT-5.4 prompted with our codebook, i.e., the scanners' model asked our questions; and our detector. We evaluate all methods on the 400-unit gold set (61 loosening directives), the only labelled set unseen by our detector, reporting precision, recall with 95\% confidence intervals, and F1 for loosening directives, and the share of each method's flags that fall on tightening directives (\%T). We compare our detector with the codebook-prompted model using McNemar's test. Because offline scanning needs no API calls, we also run the offline scanners on the full prevalence sample to estimate recall per mechanism. On the gold set, we also report the share of skills each scanner flags at all.

\begin{table}[t]
\centering
\caption{Flagging loosening directives on the 400 gold units (61 loosening). Scanner flag: a medium or higher finding on the unit's line or quoting it. LLM: GPT-5.4. \%T: share of flags on tightening directives.}
 \vspace{-4pt}
\label{tab:RQ4}
\begin{tabular}{lrrrr}\toprule
\textbf{Method}& \textbf{Prec.}& \textbf{Recall [95\% CI]}& \textbf{F1} &\textbf{\%T} \\
\midrule
Keyword lexicon & 0.20& 0.82 [0.71, 0.90]& 0.33&24\%\\
Cisco offline& 0.14& 0.02 [0.00, 0.09]& 0.03&29\%\\
Cisco + LLM& 0.39& 0.15 [0.08, 0.26]& 0.21&13\%\\
SkillSpector offline& 0.21& 0.21 [0.13, 0.33]& 0.21&29\%\\
SkillSpector + LLM& 0.27& 0.46 [0.34, 0.58]& 0.34&23\%\\
Either + LLM& 0.28& 0.51 [0.39, 0.63]& 0.36&23\%\\
\midrule
LLM + our codebook & 0.71& 0.61 [0.48, 0.72]& 0.65&4\%\\
Our detector& 0.70& 0.61 [0.48, 0.72]& 0.65&4\%\\
\bottomrule
\end{tabular}
\end{table}

\textbf{Finding 4.1: Offline scanners miss nearly all loosening directives.} On the prevalence sample, the Cisco scanner flags 1.6\% of loosening directives, all of them instructions to ignore previous instructions, and SkillSpector flags 28\%. Neither flags a single instruction not to stop or a single consequential-action directive. On the gold set, their F1 scores of 0.03 and 0.21 are below that of plain keyword matching (0.33; Table~\ref{tab:RQ4}).

\textbf{Finding 4.2: LLM analysis helps the scanners but does not close the gap.} With GPT-5.4 enabled, recall rises from 0.02 to 0.15 for Cisco and from 0.21 to 0.46 for SkillSpector. Together, the two scanners catch about half of the loosening directives (0.51), but at a precision of 0.28. Neither catches any directive that skips a safeguard, and they catch at most two of the eight instructions not to stop. SkillSpector with LLM analysis flags 90\% of gold skills whose unit is loosening, but also 79\% and 85\% of those whose unit is tightening or neither, so a flag carries little information.

\textbf{Finding 4.3: Scanners frequently mistake guardrails for risks.} Between 23\% and 29\% of the scanners' flags fall on tightening directives (13\% for Cisco with LLM analysis). SkillSpector's excessive-agency rule, for example, fires on "do not auto-approve" and "never overwrite silently" as well as on their loosening counterparts. By comparison, 4\% of our detector's flags fall on tightening directives.

\textbf{Finding 4.4: The question asked, not the model, limits the scanners.} Asked our codebook's question instead of whether a skill is malicious, the same model behind the scanners' LLM analysis reaches an F1 of 0.65, nearly twice the best scanner configuration (0.36), and statistically indistinguishable from our detector ($p = 1.0$).

\begin{tcolorbox}[colback=white, colframe=black, boxrule=0.5pt, arc=0pt, left=6pt, right=6pt, top=4pt, bottom=4pt]
\textit{\textbf{Answer to RQ4:} Offline scanners catch 2--28\% of loosening directives, and 15--46\% with LLM analysis, at a precision below 0.4; a quarter of their flags land on guardrails. The same LLM, asked an oversight-specific question, matches our detector (F1 0.65).}
\end{tcolorbox}

\section{Discussion and Implications}\label{sec:discuss}
\newcommand{\discpoint}[1]{\par\smallskip\noindent\textbf{#1}\enspace}
\newcommand{\implication}[2]{\par\vspace{1pt}\noindent\hspace{0.8em}%
  $\triangleright$\,\textit{Implication for #1.}\enspace #2\par}

\discpoint{A loosening directive is a convenience, not an attack.}
Most loosening directives make a workflow more convenient: they pass \texttt{-{}-yes} to a package manager or let a test loop run uninterrupted (Finding~1.2), and they cluster in skills that run shell commands or install packages, where waiting for a human is presumably most tedious (Findings~2.1 and~2.2). Unlike the malicious skills catalogued in prior work~\cite{registry2026malicious}, they show no sign of intent to harm, and unlike the vulnerabilities studied by Liu et al.~\cite{liu2026skills}, they work as their authors meant them to. The risk is that a file designed to be shared carries these directives to agents and contexts the author did not anticipate, without a signal to check when oversight is being skipped.
\implication{tool builders}{A flagged loosening directive should prompt a review, not a verdict.}

\discpoint{Scope is a missing safeguard.} Over two thirds of loosening directives are unscoped (Finding~1.3), setting one level of autonomy for every context a skill runs in~\cite{parasuraman2000model,feng2025levels}.

\implication{skill authors and agent vendors}{Authors can keep most of the
convenience with a one-line scope clause (``pass \texttt{--yes} only in
CI''). Vendors can warn about unscoped loosening at install time,
especially in skills that run shell commands.}

\discpoint{Fixes should be aimed at upstream skills, not their adopters.} Copying does not amplify loosening (Finding~3.1), and 92-97\% of variants leave oversight text untouched (Finding~3.3), but loosening is concentrated (Finding~3.2): 960 of 86,048 distinct loosening units (1.1\%) account for half of its occurrences. Cursor rules show the same reuse~\cite{jiang2026cursor}.
\implication{maintainers of upstream skills and catalog curators}{Editing the 100 families carrying the most loosening would reach almost a quarter of the loosening-carrying skills that agents load, and reviewing those 960 units by hand covers half of all loosening occurrences.}

\discpoint{Ask scanners a different question.}
Existing scanners treat skills as a malware problem, in their rules and in their prompts, and loosening directives contain no malware (Findings~4.1 and~4.4). 
\implication{scanner developers}{An oversight-aware check is a cheap extension: a polarity-aware rule set that tells ``never use \texttt{--force}'' from ``use \texttt{--force}'', or a dedicated oversight question to the model they already call, evaluated on labelled data such as ours.}

\discpoint{Implications for research.} Prior studies of agent instruction files catalogue what they tell agents about a project, such as how to build, test, and structure it, and find that few address security~\cite{chatlatanagulchai2025agentreadmes}; studies of skills ask whether they are malicious or vulnerable~\cite{liu2026skills,registry2026malicious}. Neither asks how instructions shift control between an agent and its human. Our work shows that instructions both add and remove oversight, and opens a line of inquiry into how agent instructions shape human oversight in both directions.

We measure what skills say, not how agents act. Future work is encouraged to test whether agents follow oversight directives and whether a scope clause changes their behaviour, improve the lexicon's recall (\S\ref{sec:threats}), extend the analysis to \texttt{AGENTS.md} and other instruction files, and track loosening over time.

\section{Threats to Validity}\label{sec:threats}
\noindent \textbf{Construct validity.} The boundary between a directive and a mention, and the polarity of directives, involve judgment. We address this with a codebook refined over three versions, independent double coding ($\kappa$ = 0.80--0.82 on the three-way label, 0.92--0.97 on polarity), and adjudication. The second coder was an LLM: no second human coder was available, LLMs can reach near-human agreement on labelling with predefined categories~\cite{ahmed2025llms}, and it made double coding of 1,000 units feasible. Its labels were committed before comparison, but agreement reflects a human and a model rather than two humans. The pre-labelled units were reviewed by the first author and may be anchored by the proposals, so we do not count them toward reliability. Two conventions shape our numbers: we count commands that an agent is told to run with prompt-suppressing flags as loosening, and we define persistence broadly, so that an instruction not to stop counts as loosening whether it skips a human check-in or only continues past a failure or until the task is complete (e.g., ``don't stop at the first failure''). Counting only the former (9 of 30 persistence units are of the latter kind), loosening prevalence is 12.7\% [11.0, 14.4] rather than 13.3\%.

All analyses start from candidate units that match our lexicon. To estimate what it misses, we drew 250 random skills (249 with text; 28,311 units) and pre-screened every unit with both detector components, RoBERTa and Jev. The first author then labelled a stratified sample of 319 units, blind to stratum and model output, and we weighted the labels back to the full sample. Among candidate units, the two classifiers flagged 94\% of the verified loosening and all of the verified tightening, so few directives should escape the screen. The lexicon captures an estimated 32\% [14\%, 75\%] of loosening directives (61\% without one unit of sampling weight 28) and 16\% [10\%, 25\%] of tightening directives. Missed phrasings include ``no approval needed'' and ``ask for confirmation before every state-changing action''. Our prevalence estimates are therefore conditional on candidate units, and absolute counts are lower bounds. Since the lexicon misses tightening more often than loosening, the 2.7:1 ratio is conservative. RQ2 assumes that misses do not depend on skill capability; our recall sample is too small to test this.

\noindent \textbf{Internal validity.} Skill-level analyses (Finding~1.5, RQ2, RQ3) rely on a detector with a loosening F1 of 0.65. On the repository-balanced frame it finds less loosening (10.3\%) and more tightening (39.3\%) than the human labels (13.3\% and 36.4\%); with recall (0.61) below precision (0.70), it predicts about 0.87 loosening units per true one, which accounts for most of the gap. We therefore base prevalence on human labels, override the detector with them where we can, treat detector-based counts as conservative for loosening, and replicate RQ2 with RoBERTa alone. The detector's errors do not inflate RQ2: on the gold set it undercounts loosening slightly more in skills with shell code (recall 0.64, precision 0.76) than in others (0.55, 0.60), so the shell association is, if anything, conservative. We decided to combine the two classifiers after observing both on the gold set. In RQ3, ancestors are inferred from first-commit dates or copy counts rather than observed, and the two ancestor rules disagree on small shifts, so we report both.

\noindent \textbf{External validity.} GitSkills covers public GitHub repositories; private and marketplace skills may differ. We analyze only English skills. LLM-mode scanning was evaluated on the 400 gold units with one model (GPT-5.4), so its confidence intervals are wide.

\section{Conclusion}\label{sec:conclude}
Agent skills shape how much control humans keep over AI agents, and they do so at scale. Across 1.53 million English skills, about one in seventeen (about 88,000) contains a directive that loosens that control, most often through permission flags, skipped confirmation, or instructions not to stop, and over two thirds do so unconditionally. Loosening concentrates in skills that can execute code and spreads through a few widely copied upstream skills rather than through users weakening their copies. Existing scanners miss most of it, not because their models cannot recognize it, but because they are not asked to look for it. We believe our codebook, labels, and detector would help skill authors, users, and scanner developers weigh convenience against oversight.

\bibliographystyle{ACM-Reference-Format}
\bibliography{references}

\end{document}